\documentclass[intlimits,twoside,a4paper]{article}

\usepackage{bbold}                 %%%% package for black-bord lowercase letters
\usepackage{multirow}
\usepackage[cp1251]{inputenc}

\usepackage[eqsecnum]{cmpj3}

\articletype{A part of the Special collection to the 100th anniversary of the birth of Ihor Yukhnovskii}

\issue{2025}{28}{2}{23801}
\doinumber{10.5488/CMP.28.23801}

\title[Excluded volume effect of surfactant ligands on the shape of nascent nanocrystal]
{Excluded volume effect of surfactant ligands on the shape of nascent nanocrystal}

\author[A. Baumketner, D. Anokhin, Ya. Patsahan]{A. Baumketner\orcid{0000-0003-2726-931X}\refaddr{label1}\thanks{2024/2025 Fellow of the Virtual Ukraine Institute for Advanced Study (VUIAS)}~\thanks{Corresponding author: \email{andrij@icmp.lviv.ua}.}, 
    D. Anokhin\orcid{0000-0002-4958-2692}\refaddr{label2}, Ya. Patsahan\orcid{0009-0000-2593-027X}\refaddr{label3}}
\addresses{
\addr{label1} Institute for Condensed Matter Physics of the National
Academy of Sciences of Ukraine, 1 Svientsitskii St., 79011 Lviv,
Ukraine
\addr{label2} School of Chemistry, V. N. Karazin Kharkiv National University, 
4 Svobody Sq., 61022 Kharkiv,  Ukraine
\addr{label3} Department of Physics,  Ivan Franko National University of  Lviv,
 8a Kyrylo and Mefodiy St., 79005 Lviv, Ukraine
}

\Keywords{nanocrystals, hard-sphere model, nanoplates, nanorods,  Monte Carlo simulation}

\date{Received April 7, 2025, in final form May 31, 2025}

\begin{document}

\maketitle

\begin{abstract}
We investigate the effect of the excluded volume of surfactant ligands on the shape of incipient quantum dots (QDs) to which they are attached.  We consider a model in which ligands are represented by hard-sphere particles that are bound to the surface of a nanoparticle (NC) that is cast in the shape of a prism. It is found in Monte Carlo simulations that the ensemble of relevant NC conformations consists of a small number of specific states that take on the form of nanoplates and nanorods. The shape of these states can be well described by the derived theoretical models. At increasing ligand density, the free energy of different states is seen  to be approximately the same, suggesting that excluded volume interactions among ligands acts to narrow down the conformational space accessible to an NC without creating a statistical preference for any particular configuration.
%
%
%\keywords Up to six keywords (\href{https://physh.aps.org/browse}{Physics Subject Headings})
\printkeywords
\end{abstract}

\section{Introduction}

Quantum dots (QD) are colloidal particles of nanometer dimensions, which --- due to the quantum confinement effect-exhibit unique and size-dependent optical and electronic properties. Lead halide perovskite (LHP) nanocrystals (NCs) is the latest addition to the family of QDs designed specifically for optical applications such as light generation or detection. Aided by unique chemistry and facile and inexpensive synthesis, these materials rely on perovskite structure to achieve superior narrowband photoluminescence (PL) with near-unity quantum yield (QY) and wide spectral tunability (410--750 nm), high absorption coefficients and high radiative rates and long excitonic coherence times~\cite{Rai16}.

Most commonly quantum dots are cubic~\cite{Prot15}, although a variety of alternative shapes have been reported, including nanocubes~\cite{Prot15}, nanoplatelets~\cite{Pan16}, truncated octahedra~\cite{Zhang19}, spherical colloids~\cite{Zhang19,Akk22} and nanorods~\cite{Wang19,Zhu22}. Notably, different shapes produce different quantum confinements --- 3D, 2D and so on, thus directly affecting various optical properties including anisotropic and polarized emission~\cite{Scott17,Hik05}, width of the emission lines~\cite{Lhu15}, size of the absorption cross-section~\cite{Yel15} and amplified spontaneous emission thresholds~\cite{Dede19, Kel19}.

Nucleation and growth of QDs represent a unique example of the self-assembly process whose mechanistic details, including factors controlling geometry and relevant growth characteristics,  are of primary interest to fundamental research in physics and chemistry. Furthermore, a microscopic understanding of how QDs are formed can open the door to a rational design of NCs with preset geometries, which can be desirable under a variety of conditions. For instance, it is known that cubic NCs have a degenerate excited state causing a splitting in the associated emission line due to the symmetry imposed by all three sides of the nanoparticle having the same length. The splitting is detrimental because it reduces the purity of the emission line. As a potential remedy, it is very tempting to suggest growing asymmetrical particles, for instance nanorods, which have a reduced symmetry and thus may have an improved spectrum. Naturally, one needs to know how to design non-cubic nanocrystals in order to test this assumption, which is a goal that requires an in-depth understanding of the microscopic factors controlling the shape of the NCs.

Surfactant ligands are essential for the entire life cycle of  LHP NCs, starting from  nucleation/growth phase and ending with the thermal stability~\cite{Prot15}. They are also responsible, to a large degree, for defining the  shapes of QDs. Different ligands lead to different shapes suggesting that ligand-ligand interactions play a critical role. Although ligands are formed in a variety of types depending on their chemistry and length, there is one characteristic feature they have in common  --- a finite size. Thanks to the finite size  no two ligands can occupy the same volume in space, leading to the  arisal of ligand-ligand repulsion. Here, we focus on this so-called excluded volume effect --- in its simplest representation --- in the context of a growing nanocrystal. 

We consider a model in which ligands are treated as hard-sphere particles whose centers are bound to the surface of a prism --- a geometrical figure that has a square base and variable height. 
We note that
this description was chosen on the grounds of simplicity and disregards a whole variety of technical
details, such as non-spherical shape of the ligand, ligand-ligand interactions of finite range, the ability
of the ligands to detach from the surface, anisotropic binding affinities, or surface energy effects, which
can be included at a later time.
The prism can capture both nanoplatelet and nanorod shapes of the quantum dots. Configurations of spheres on prism facets are studied by Monte Carlo simulations. As the sphere density is increased, various geometries of the prism begin to
diverge in terms of their observation probabilities.
The free energy of these geometries is evaluated by performing a thermodynamic integration over pressure. At a certain high density, the entire ensemble of NC configurations is partitioned into a series of allowed and forbidden configurations, where the forbidden configurations correspond to  free energy barriers while  the allowed ones --- to free energy minima. The allowed  configurations are partitioned  into  the
nanoplatelet and nanorod families. Interestingly, at the point where spontaneous transitions between allowed  states become forbidden, the ligand sub-system still remains in a fluid state at a low density, suggesting that no phase transition or  even  kinetic slowdown takes place. We conclude based on these findings that ligands can considerably influence the shape of a growing NC even at a low surface coverage purely through an excluded volume effect.  This circumstance needs to be taken into account when designing novel ligands for the synthesis of NCs with specific properties.
\begin{figure}[!t]
\centerline{\includegraphics[scale=0.4]{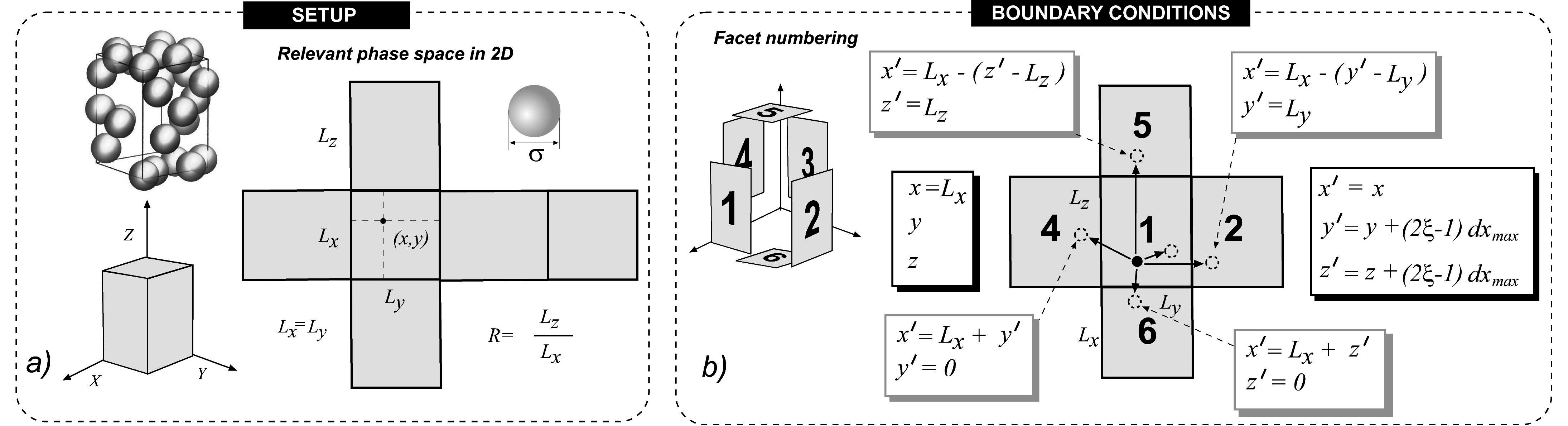}}
\caption{Details of the model studied. (a) A number of 3D hard-sphere particles representing ligands are distributed on the surface of a prism. Particles can move across all facets, thus exploring the depicted  2D phase space. (b) Transitions between facets are governed by specific boundary conditions. A particle in facet 1 has 5 different options for a translational move as explained in the text. The boundary conditions are specific to each facet.}
\label{fig1}
\end{figure}
\section{Model and methods}
We consider ligands to be hard-sphere particles of diameter $\sigma$ lying on the surface of a prism. A prism is a solid three-dimensional  object with two identical bases, see figure~\ref{fig1}(a) for illustration. Depending on the type of the polygon making up the base, the prisms can be classified as triangular, square, rectangular and so on. We are concerned with the square prism, which can be either a rod or a platelet, depending on the aspect ratio of its sides. 

Each particle is a sphere of diameter $\sigma$ characterized by two independent coordinates whose identity depends on the facet. All facets are numbered as explained in figure~\ref{fig1}(b). For facets 5 and 6, the respective coordinates are $x$ and $y$. For facets 1 and 3 --- $y$ and $z$, and for facets 2 and 4 --- $x$ and $z$. 
Although ligand particles are treated as 3D objects, the effective space in which they are allowed to move is two-dimensional, as it is made by  joining 6 separate facets together, see figure~\ref{fig1}(a) for an explanation. 

\subsection{Fixed cell geometry}
The coordinates are advanced on the 2D plane according to the rules specific to each facet. For instance, in the case of facet 1, the relevant coordinates are $y$ and $z$ and trial coordinates in a Monte Carlo (MC) move are  generated as: 

\begin{equation}
\begin{aligned}
    &x'=x=L_x, \\
    &y'=y+(2\xi-1)d x_{\text{max}},\\
    &z'=z+(2\xi-1)d x_{\text{max}},
\end{aligned}
\label{eq1}
\end{equation}
\begin{table}[!t]
	\caption{Explanation of the rules used to generate trial positions in Monte Carlo moves. As the first step, random displacements are applied to the two coordinates that are considered variable, according to the facet. If the generated displacement is seen to take the particle outside of its current facet, appropriate boundary conditions are applied, enabling facet-to-facet transitions. The corresponding transformations of the coordinates are listed separately for each facet. \textbf{$\xi$} is a random variable between 0 and 1. 
		%\textbf{{$\textit{\textbf{dx}}_{\text{max}}$}} 
		$dx_{\text{max}}$ is an adjustable parameter. Other notations are as in the main text.}
	\vspace{0.3cm}
\centering
\begin{tabular}{ccc|c}

\hline
                        & \multirow{6}{*}{\includegraphics[scale=0.6]{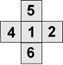}}    & 1\textrightarrow2  & 1\textrightarrow4 \\
\textbf{Facet 1}        &                                                                    & $x'=L_x-(y'-L_y)$  & $x'=L_x+y'$       \\
$x'=L_x$                &                                                                    & $y'=L_y$           & $y'= 0$           \\ \cline{3-4} 
$y'=y+(2\xi-1)dx_{\text{max}}$ &                                                                    & 1\textrightarrow5  & 1\textrightarrow6 \\
$z'=z+(2\xi-1)dx_{\text{max}}$ &                                                                    & $x'=L_x-(z'-L_z)$  & $x'=L_z+z'$       \\
                        &                                                                    & $z'=L_z$           & $z'=0$            \\ 
\hline
                        & \multirow{6}{*}{\includegraphics[scale=0.6]{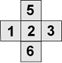}}    & 2\textrightarrow3  & 2\textrightarrow1 \\
\textbf{Facet 2}        &                                                                    & $y'=L_y+x$         & $y'=L_y-(x'-L_x)$ \\
$x'=x+(2\xi-1)dx_{\text{max}}$ &                                                                    & $x'=0$             & $x'= L_x$         \\ \cline{3-4} 
$y'=y+L_y$              &                                                                    & 2\textrightarrow5  & 2\textrightarrow6 \\
$z'=z+(2\xi-1)dx_{\text{max}}$ &                                                                    & $y'=L_y-(z'-L_z)$  & $y'=L_y+z'$       \\
                        &                                                                    & $z'=L_z$           & $z'=0$            \\ 
\hline
                        & \multirow{6}{*}{\includegraphics[scale=0.6]{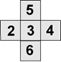}}    & 3\textrightarrow4  & 3\textrightarrow2 \\
\textbf{Facet 3}        &                                                                    & $x'=-y$            & $x'=y'-L_y$       \\
$x'=0$                  &                                                                    & $y'=0$             & $y'= L_y$         \\ \cline{3-4} 
$y'=y+(2\xi-1)dx_{\text{max}}$ &                                                                    & 3\textrightarrow5  & 3\textrightarrow6 \\
$z'=z+(2\xi-1)dx_{\text{max}}$ &                                                                    & $x'=z'-L_z$        & $x'=-z'$          \\
                        &                                                                    & $z'=L_z$           & $z'=0$            \\ 
\hline
                        & \multirow{6}{*}{\includegraphics[scale=0.6]{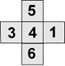}}    & 4\textrightarrow1  & 4\textrightarrow3 \\
\textbf{Facet 4}        &                                                                    & $y'=x'-L_x$        & $y'=-x'$          \\
$x'=x+(2\xi-1)dx_{\text{max}}$ &                                                                    & $x'=L_x$           & $x'=0$            \\ \cline{3-4} 
$y'=0$                  &                                                                    & 4\textrightarrow5  & 4\textrightarrow6 \\
$z'=z+(2\xi-1)dx_{\text{max}}$ &                                                                    & $y'=z'-L_z$        & $y'=-z'$          \\
                        &                                                                    & $z'=L_z$           & $z'=0$            \\ 
\hline
                        & \multirow{6}{*}{\includegraphics[scale=0.6]{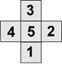}}    & 5\textrightarrow2  & 5\textrightarrow4 \\
\textbf{Facet 5}        &                                                                    & $z'=L_z-(y'-L_y)$  & $z'=L_z+y'$       \\
$x'=x+(2\xi-1)dx_{\text{max}}$ &                                                                    & $y'=L_y$           & $y'=0$            \\ \cline{3-4} 
$y'=y+(2\xi-1)dx_{\text{max}}$ &                                                                    & 5\textrightarrow5  & 5\textrightarrow1 \\
$z'=L_z$                &                                                                    & $z'=L_z+x$         & $z'=L_z-(x'-L_z)$ \\
                        &                                                                    & $x'=0$             & $x'=L_x$          \\ 
\hline
                        & \multirow{6}{*}{\includegraphics[scale=0.6]{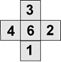}}    & 5\textrightarrow2  & 5\textrightarrow4 \\
\textbf{Facet 6}        &                                                                    & $z'=y'-L-y$        & $z'=-y'$          \\
$x'=x+(2\xi-1)dx_{\text{max}}$ &                                                                    & $y'=L_y$           & $y'=0$            \\ \cline{3-4} 
$y'=y+(2\xi-1)dx_{\text{max}}$ &                                                                    & 5\textrightarrow5  & 5\textrightarrow1 \\
$z'=0$                  &                                                                    & $z'=-x'$           & $z'=x'-L_x$       \\
                        &                                                                    & $x'=0$             & $x'=L_x$          \\ 
\hline
\end{tabular}
\label{table1}
\end{table}
where $x,y,z$ are the current coordinates, $\xi$ is a random number uniformly distributed between 0 and 1, $dx_{\text{max}}$ is the maximum displacement and $L_x$ is the length of the facet in $x$ direction. Equivalent quantities for $y$ and $z$ dimensions are  $L_y$ and  $L_z$. The aspect ratio of the simulation cell is defined as $R=L_z/L_x$.

Trial coordinates are accepted if they do not cause any steric clashes among particles and rejected otherwise. If the trial coordinates fall outside the ranges set by the box geometry,  $0<x'<L_x$, $0<y'<L_y$ and $0<z'<L_z$, transitions between facets are arranged. For facet number 1 for instance, particles can exit through the border with facet number 2, in which case the identity of the particle coordinates changes. To reflect this change, the coordinates are reassigned according to geometrical considerations explained in figure~\ref{fig1}(b). Namely, $z'$ coordinate remains unchanged while  $x'$ is assigned a new value  $L_x-(y'-L_y)$ and $y'$ is assigned a new value $L_y$, in that particular order. As a result of this transformation, the particle in question jumps from facet 1 to facet 2 and then continues moving on the new facet. For transitions into other facets --- 5 , 4 and 6, similar formulas are employed as explained in figure~\ref{fig1}(b). Similar formulas are derived for facets 2--6, leading to a complete set of rules shown in table~\ref{table1}. By design, the rules ensure that the hard-sphere particles perform a random and unbiased walk on the surface of a prism. For a single particle, the probability of being observed on a particular facet is then proportional to the area of its surface. This law was tested and verified in explicit simulations. Moves in which particles exit a facet through the corner were disregarded because they can be represented by a sequence of 2 moves in which particles exit through the side. Similarly, moves in which particles jump the length of an entire facet were also avoided  (by adjustment of parameter $dx_{\text{max}}$).

\subsection{Volume moves}

Density of ligands on the surface of a nanocrystal, or the surface coverage, can be varied by tuning the conditions of the experiment. In the studied model we consider pressure as an external parameter that controls the ligand density. Increasing pressure leads to a higher density while an opposite effect is seen for a decreasing pressure. A constant-pressure constant-temperature (CPT) algorithm relying on the scaling of the size of the simulation box~\cite{Allen86} is used to maintain steady pressure in our simulations. At the heart of the algorithm are volume moves designed to steer the system into a state consistent with the applied pressure $P$. This is achieved by introducing a set of scaled coordinates that relate actual coordinates with the volume of the relevant phase space. In the case of 2D system, volume is replaced by the surface area. If $(\varmathbb{x}_i, \varmathbb{y}_i), i=1,N$, are coordinates of some $N$-particle system that occupies surface area $S$, the scaled coordinates are introduced as:
\begin{equation}
\begin{aligned}
    &\varmathbb{x}_i=\sqrt{S}\varmathbb{x}_i^S \\
    &\varmathbb{y}_i=\sqrt{S}\varmathbb{y}_i^S.
\end{aligned}
\label{eq2}
\end{equation}

Thermodynamic expectation of some configurational variable $A(\varmathbb{x}_1, \varmathbb{y}_1 \dotsc\varmathbb{x}_N, \varmathbb{y}_N)$ at pressure $P$ can then be written as: 
\begin{align}
    \langle A \rangle_{CPT} &= \frac{\int \rd S \re^{-\beta PS}\int \re^{-\beta U}A\rd \varmathbb{x}_1 \rd \varmathbb{y}_1\dotsc \rd \varmathbb{x}_{N}\rd\varmathbb{y}_N}{\int \rd S \re^{-\beta PS}\int \re^{-\beta U}\rd\varmathbb{x}_1\rd\varmathbb{y}_1\dotsc \rd\varmathbb{x}_N\rd \varmathbb{y}_N}\nonumber\\
    &=\frac{\int \rd S \re^{-\beta PS}\int \re^{-\beta U}AS^N \rd \varmathbb{x}^S_1 \rd\varmathbb{y}^S_1\dotsc \rd\varmathbb{x}^S_N \rd\varmathbb{y}^S_N}{\int \rd S \re^{-\beta PS}\int \re^{-\beta U}S^N \rd\varmathbb{x}^S_1\rd \varmathbb{y}^S_1\dotsc \rd\varmathbb{x}^S_N\rd\varmathbb{y}^S_N},
\label{eq3}
\end{align}
where $U$ is potential energy and we transitioned to the set of scaled variables on the right-hand side. It can be seen that the average $\langle A \rangle_{CPT}$ can be obtained by sampling in the combined set of variables $(S, \varmathbb{x}^S_1\varmathbb{y}^S_1\dotsc \varmathbb{x}^S_N\varmathbb{y}^S_N)$ performed with limiting distribution function $\re^{-\beta(PS+U-kTN\log(S))}$,where $\beta=1/kT$ and $k$ is the Boltzmann's constant and $T$ is the temperature. In practical terms, a volume move entails an attempt to change the current surface area $S$ to a trial value $S'$. If the importance sampling algorithm is used, the probability of making such a transition is governed by the function $\re^w=\re^{-\beta(P\Delta S+\Delta U-kTN\log(\frac{S'}{S}))}$, where $\Delta S=S'-S$ and $\Delta U=U(\sqrt{S'}\varmathbb{x}^S_i,\sqrt{S'}\varmathbb{y}^S_i)-U(\sqrt{S}\varmathbb{x}^S_i,\sqrt{S}\varmathbb{y}^S_i)=0$ for hard-sphere system provided that new volume is allowed by steric constraints. Taking into account relations (\ref{eq2}) it is easy to see that $\sqrt{\frac{S'}{S}}=\frac{\varmathbb{x}'_i}{\varmathbb{x}_i}=\frac{\varmathbb{y}'_i}{\varmathbb{y}_i}=\eta$, where $\varmathbb{x}'_i$, $\varmathbb{y}'_i$ are trial coordinates resulting from a volume move and $\eta$ is the scaling factor by which the coordinates  are multiplied. With these notations, the move-acceptance function can be re-written as $w=-\beta P\Delta S+N_f\log(\eta)$, where $N_f=2N$ is the number of degrees of freedom that get scaled during a move.

As a practical implementation of this algorithm, we generate trial coordinates by scaling the current coordinates by a random scaling factor $\eta$ close to unity:

\begin{equation}
\begin{array}{ccc}
    x'_i=\eta x_i, &          & L'_x=\eta L_x ,\\
    y'_i=\eta y_i, & i=1,N,    & L'_y=\eta L_y, \\
    z'_i=\eta z_i, &          & L'_z=\eta L_z.
\end{array}
\label{eq4}
\end{equation}

The surface area $S=2L^2_x+4L_xL_z$ then changes to $S'=\eta^2(2L^2_x+4L_xL_z)$ so that $\Delta S= (\eta^2-1)S$ or, if we invert that relation, $\eta=\sqrt{1+\frac{\Delta S}{S}}$. In order to evaluate integral (\ref{eq3}) by stochastic MC method, surface area $S$ should be treated as a \textit{\textbf{uniformly distributed}} random variable, implying that $\Delta S$  is a variable of the same kind. Hence, the rule for generating new dimensions of the cell becomes

\begin{equation}
\begin{array}{ccc}
    L'_j=\eta L_j, & j=x,y,z,    & \eta=\sqrt{1+\frac{(2\xi-1)\Delta S_{\text{max}}}{S}}. 
\end{array}
\label{eq5}
\end{equation}

Where $\xi$, as before, is a random number uniformly distributed between 0 and 1, and $\Delta S_{\text{max}}$ is an adjustable parameter. Trial coordinates are generated from $\eta$ by equation (\ref{eq4}) while the trial change of surface area is computed as $\Delta S=(\eta^2-1)S$. The test function

\begin{equation}
    w = -\beta P\Delta S+N_f\log(\eta)
\label{eq6}
\end{equation}
is evaluated to determine whether to accept --- $\re^w\geqslant\xi$ --- or reject --- $\re^w\leqslant\xi$ --- the trial move, where $\xi$ is a random number between 0 and 1.

It follows from (\ref{eq4}) that the aspect ratio of the cell $R'=\frac{L'_z}{L'_x}=\frac{L_z}{L_x}=R$ is preserved under the volume-change moves. This feature may or may not be desirable depending on the purposes. If the goal is to evaluate relative free energy of cells with different $R$'s, then the derived algorithm is appropriate. However, if the goal is to locate most likely geometries under a given pressure, then the volume-move algorithm must allow $R$ to vary. There are multiple ways of achieving this, the simplest being described below. Focusing on the expression for the surface area $S=2L^2_x+4L_xL_z$ it is easy to see that $S$ can be also changed by scaling $z$-coordinates while keeping the dimensions in $x$ and $y$ directions unchanged. Repeating the steps taken above for the other volume-change move, this will lead to the following algorithm
\begin{equation}
\begin{array}{ccc}
    x'_i=x_i,      &          & L'_x=L_x, \\
    y'_i=y_i ,     & i=1,N,    & L'_y=L_y, \\
    z'_i=\eta z_i, &          & L'_z=\eta L_z.
\end{array}
\label{eq7}
\end{equation}
The corresponding surface area change $\Delta S=S'-S=2L_x^2+\eta 4L_xL_z-2L_x^2-4L_xL_z=(\eta-1)4L_xL_z$ leads to the expression of the scaling coefficient $\eta=1+\frac{\Delta S}{4L_xL_z}$. The trial geometry is then generated as
\begin{equation}
    L'_z=\eta L_z, \quad L'_x=L_x, \quad L'_y=L_y, \quad \eta=1+\frac{(2\xi-1)\Delta S^z_{\text{max}}}{4L_xL_z},
\label{eq8}
\end{equation}
where $\xi$ is the random variable introduced earlier and $\Delta S^z_{\text{max}}$ is an adjustable parameter (not to be confused with $\Delta S_{\text{max}}$). The moves are accepted or rejected by evaluating the test function $w$ and following the rules outlined in formula (\ref{eq6}). Compared to the previous algorithm,  however, there is now an important difference in how $N_f$ is evaluated. While previously the scaling applied to both coordinates of all particles, at present it applies only to $z$-dimension, thus affecting only the particles located on the facets 1 through 4, see figure~\ref{fig1}(b). Accordingly, the number of involved degrees of freedom is $N_f=N_{\text{side}}$, where $N_{\text{side}}$ is the number of particles on the side facets. This number needs to be evaluated before each $z$-volume change step. In other respects, the two volume-change algorithms are identical.

We performed two types of constant-pressure simulations as part of this project. First, these were simulations following the algorithm (\ref{eq4})--(\ref{eq5}) and preserving the aspect ratio of the cell. Second, these were simulations combining algorithms (\ref{eq4})--(\ref{eq5}) with those of (\ref{eq7})--(\ref{eq8}) in equal proportion. The latter allowed the simulation cell to relax, thus permitting observation of the most likely geometry. Parameters  $\Delta S_{\text{max}}^z$ and $\Delta S_{\text{max}}$ were adjusted to achieve $>50$\% success rate of the volume moves. Volume moves were attempted every 10th particle move, on average. The maximum displacement for transitional moves $dx_{\text{max}}$ was adjusted to achieve the move acceptance rates better than 50\%.

\subsection{Free energy calculation}
To calculate the Gibbs free energy for a system with fixed aspect ratio $R$, simulations are performed using algorithm (\ref{eq4})--(\ref{eq5}) at a sequence of pressures ranging from $P_0$ to $P$, where $P$ is the target pressure. Pressures are reported in dimensionless units $P^*=\beta P\sigma^2$ while the particle density --- in terms of packing fraction $\rho^*=\eta_p=\frac{\piup\sigma^2}{4}\frac{N}{S}$. The lower limit $P_0$ was selected on the condition that there should be no variation with $R$ in the average surface area observed at that pressure --- a consequence of the system at extremely low dilution having no preference for a particular geometry. By systematic analysis we found that  $P_0^*=0.12$ meets this criterion. A number of points with increasing pressure $P>P_0$ were then  considered, starting with $P_0$ and ending at 1.44 with the increment of 0.12, to  yield $S(P,R)$ --- the dependence of surface area on pressure. We observed a strong variation with $P$ in this initial segment of $S(P,R)$. It  was followed by a second segment, stretching from $P^*=1.44$ and ending at $P^*=5.29$, in which the decline was much more leveled  and which was evaluated with a larger increment of 0.48. The combined curve was then integrated numerically to yield the dependence of Gibbs free energy on $R$: 

\begin{equation}
    G(P,R)=\int^P_{P_0} S(P,R)\rd P.
\label{eq9}
\end{equation}

Since $S(P_0,R)=S(P_0)$, the integration produces a function that reports relative free energy for different values of $R$, i.e., for different cell geometries. Simpson rule~\cite{Dev93} was used to carry out the integration. By performing a comparative analytical integration on a function very closely resembling the function obtained in simulation, we estimate that the numerical integration error in the latter did not exceed 5 percentage points, which we consider to be a satisfactory accuracy.  

\section{Results}
\subsection{Suppressed elongation at low pressure}

%\doclicenseThis

\begin{figure}[htb]
\centerline{\includegraphics[width=1\textwidth]{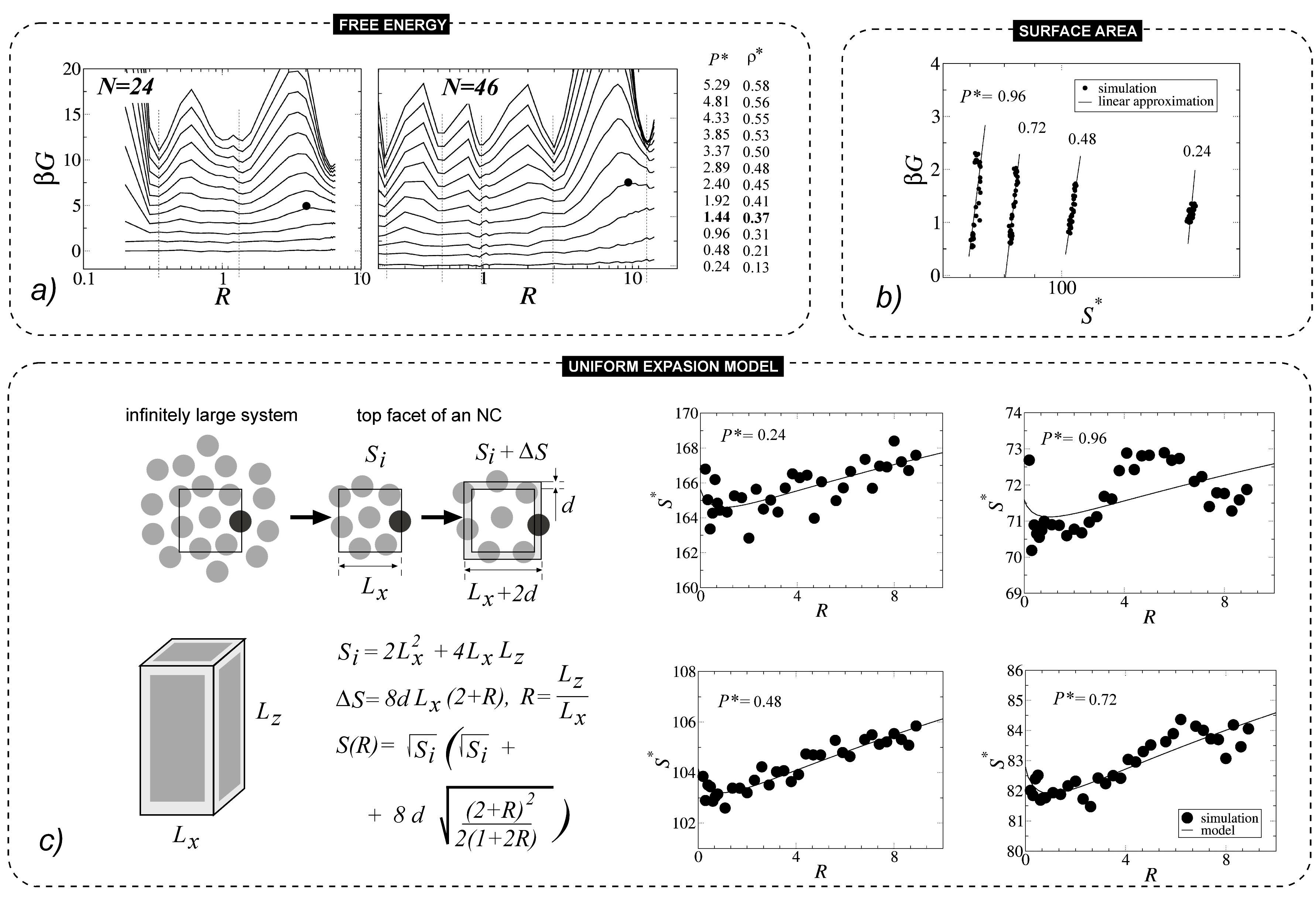}}
\caption{Various properties computed in this work for a system of $N$ hard spheres confined to the surface of a prism. (a) Gibbs free energy in dimensionless units as a function of aspect ratio $R$  computed in systems with $N=24$ and $N=46$. Broken lines mark the positions of the minima. Different curves correspond to different pressures $P^{*}=\beta P \sigma^{2}$ and average ligand densities $\rho^{*}=\eta_{p}=\frac{\piup \sigma^{2}}{4}\frac{N}{S}$, as labeled. Black circle corresponds to the pressure $P^{*}=1.44$ at which the first maximum appears. (b) Free energy vs. average surface area computed for varying $R$ in the system with $N=28$ particles and four different pressures, as labeled. Good correlation between the two properties is seen at $P^{*}=0.72$ (c) Details of the model we use to explain why rods with large $R$ have a lower probability than other shapes of the nanoparticle. Good agreement is seen for the average surface area as a function of $R$ between the model and simulation at low pressures.} \label{fig2}
\end{figure}

We computed free energy profiles $G(R)$,  as a function of the aspect ratio using the constant-pressure algorithm (\ref{eq4})--(\ref{eq5}) as described in the ``Methods'' section. The number of particles in the performed simulations varied from $N=14$ to $N=82$ so as to probe the influence of the size of the system. The results are shown in figure~\ref{fig2}(a) for a sequence of pressures ranging from $P^{*} = 0.24$ to $P^{*} = 5.29$ with an increment of 0.24. It is seen that at low pressures $P^{*}<P_{c}^{*}\approx1$, the profile is nearly flat, indicating that all cell geometries have the same statistical weight, which is expected in the low density / pressure limit. A comparison between the results obtained for $N=24$ and $N=46$ demonstrates that this behavior is consistent across  multiple system sizes.

As the pressure rises, two trends become apparent. First, the free energy at $R>1$ corresponding to nanorod geometries begins to shift upwards. Again, this happens for all system sizes. Second, at $P^{*}=1.44$, a maximum emerges in the profile. The position of the maximum shifts toward larger aspect ratios as the size of the system increases.  For $N=24$, it is seen at $R=4.3$ and for $N=46$ --- at $R=9.9$. It is interesting to explore the microscopic origins of the large-$R$ rise in $G(R)$. 

It is clear from (\ref{eq9}) that the free energy at a given pressure $P_{t}$ depends on all preceding pressures $P<P_{t}$. It is also clear that a larger surface area at the given $P_{t}$ will contribute toward larger $G(R)$ since at $P=P_{0}$, the surface area is the same for all $R$. It is possible, therefore, that $G(R)$ could be correlated with $S(R)$ within some range of pressures, in which case the latter can be used as an easier to analyze substitute for the former. Figure~\ref{fig2}(b) plots free energy vs. average surface area (shown in dimensionless units $S^{*}=S/\sigma^{2}$) observed in simulations of $N=28$ particle system with fixed cell geometry. Each point in the graph corresponds to a distinct $R$. Broken lines represent the best linear approximation of the simulation data. A very strong correlation between $S(R)$ and $G(R)$ is observed for pressures $P^{*}=0.72$ and lower, where the data correspond to a straight line. Starting with $P^{*}=0.96$, some deviations from the line begin to emerge, especially in the limit of large $G(R)$. We conclude from this analysis that average surface area can be used to explain the behavior of free energy in the limit of low pressures with the cut-off point of $P^{*}=0.72$. This is the range of pressures that is of prime interest to us --- the one where a rise in $G(R)$ is already present but there is still no maximum, see figure~\ref{fig2}(a).

How can one explain the variation of the surface area with $R$? To answer this question let us consider a system of 2D disks on a plane, which is the large-size limit of the system studied. Let us focus on a certain tagged particle that is surrounded by other particles, see figure~\ref{fig1}(c), particle marked with dark color. In the infinite system, the other particles exert pressure on the tagged one on all sides. When the tagged particle is placed near the border in a finite-size prism, the pressure on one side vanishes, causing the tagged particle to move closer to the border. This movement occurs on all borders of the prism, causing its surface area to increase. For instance, if a certain square cluster of particles carved out of the 2D plane of disks at a certain pressure $P$ occupies the surface area $S_{i}$, its size increases to $S_{i}+\Delta S$ when placed on the facets of a prism at the same pressure, where $\Delta S$ is a positive quantity. Therefore, the surface area of finite system $S(R)$ will always be greater that the corresponding surface area of infinite system $S_{i}$ for all geometries of the cell. A simple heuristic model  can be used to precisely estimate the expansion of the cell. Assume that the dimensions of the cell are $L_{x}$ and $L_{z}$  so that $S_{i}=2L_{x}^{2}+4L_{x}L_{z}$. Assume further that the expansion occurs uniformly on all sides by the same amount $d$. The surface area increase $\Delta S=4d(2L_{x}+2L_{z})+2 \times 4dL_{x}=8dL_{x}(2+R)$ can then be expressed in terms of initial size $L_{x}$ and aspect ratio $R=L_{z}/L_{x}$. Now, the initial size can be eliminated with the help of the initial surface area, $L_{x}=\sqrt{\frac{S_{i}}{2(1+2R)}}$, resulting in the following expression for the surface area of finite system:

\begin{align}
S(R)=\sqrt{S_{i}}\Bigl(\sqrt{S_{i}}+8d\sqrt{\frac{(2+R)^{2}}{2+(1+2R)}}\Bigr).
\label{eq10}
\end{align}

We treated $S_{i}$ and $d$ as fitting parameters to see how well the predicted ansatz works compared to simulation. The results are shown in figure~\ref{fig2}(c). It is seen that equation (\ref{eq10}) agrees quite well with the simulation at low pressures, where it correctly reproduces essential functional features, including a minimum at $R=1$. Deviations begin at $P^{*}=0.96$, where, as we mentioned earlier, the average surface area is no longer a good descriptive parameter for the free energy. 
Our conclusion based on this analysis is that higher free energy of nanoparticles with elongated shapes at low pressures can be explained by the general effect associated with their finite size. All finite-size systems expand in size but those with elongated shapes do so to a greater extent than others due to the specific features of their geometry. As a consequence, nanorods will appear as a slightly less prominent form of nanopartciles at a low ligand coverage. 

\subsection{Geometries allowed by excluded volume interactions}

\begin{figure}[!t]
\centerline{\includegraphics[width=1\textwidth]{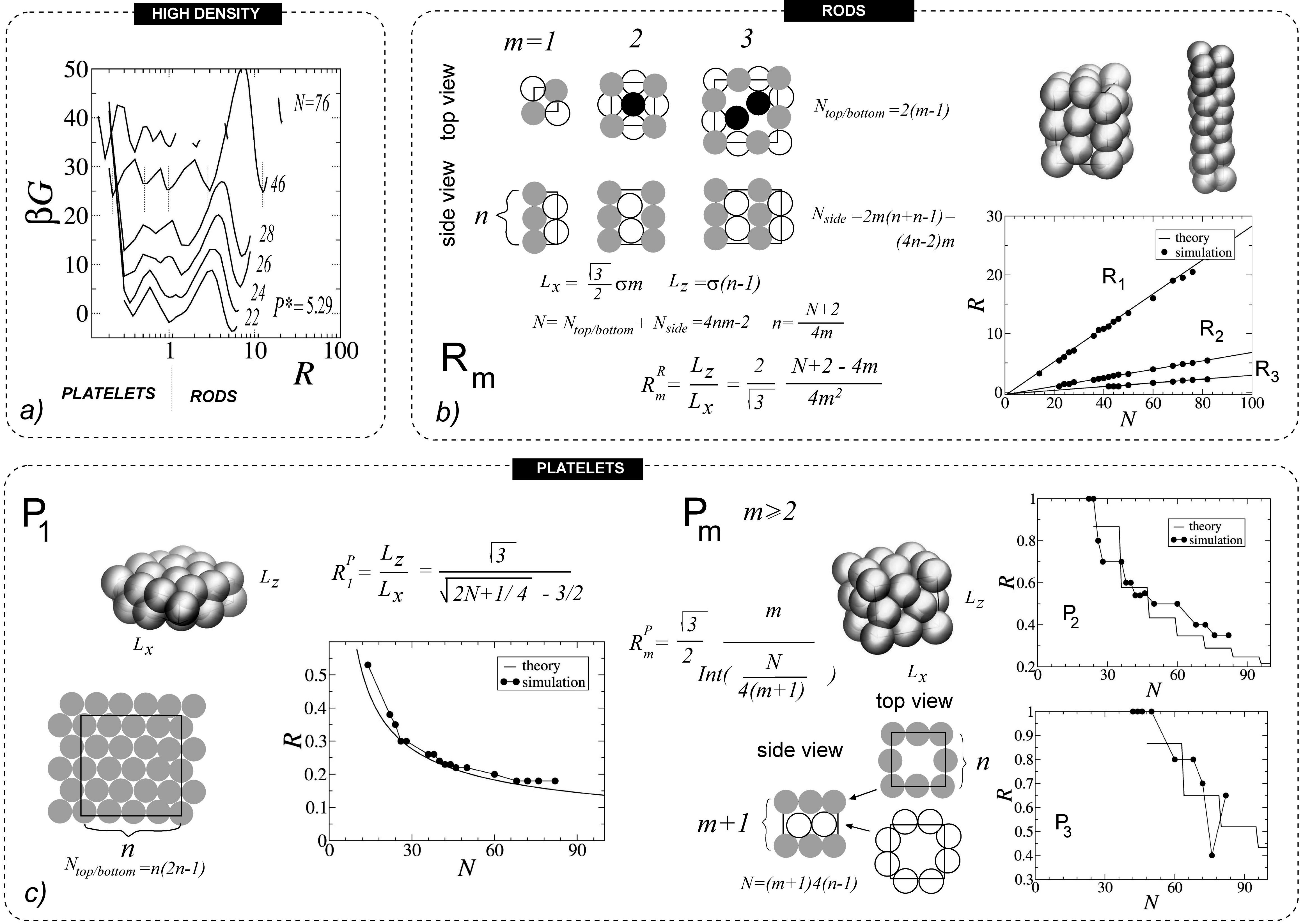}}
\caption{Various characteristics of the conformational ensemble observed for the nanoparticles at high pressure. (a) Free energy as a function of $R$ for varying $N$, as labeled. The position of  the minima is marked by a vertical broken line.  Conformations are classified as either rod or platelet. (b) Theoretical model for the structure of nanorods. Various details are as labeled and explained in the text. Comparison of the theoretical and simulation results for the aspect ratio $R(N)$ in three conformations $\textbf{R}_{1}$, $\textbf{R}_{2}$ and  $\textbf{R}_{3}$. (c) Same information as in (b) but for platelets. Simulation results are compared with theoretical predictions for $R$ for three platelet conformations  $\textbf{P}_{1}$ ,  $\textbf{P}_{2}$ and $\textbf{P}_{3}$ . Good agreement between theory and simulation is seen for all geometries.} 
\label{fig3}
\end{figure}

As pressure rises,  two changes that take place in the free energy profiles are observed. First, minima begin do develop at specific aspect ratios $R$ separated by maxima. According to figure~\ref{fig2}(a) this happens at $P^* \approx 2$ for both small system with $N=24$ and large system with $N=46$. At  $P^* \approx 5$, the minima are fully developed while the maxima have height  greater than $5kT$, which makes them essentially insurmountable for a spontaneous process driven by temperature. At this point, all conformations  of the nanoparticle can be categorized as those belonging to two zones: a) the allowed zone --- corresponding to $R$'s near the minima, and the forbidden zone  --- comprising conformations centered around maxima. The ``zone'' structure of the conformational ensemble is universal and applies to all system sizes. Figure~\ref{fig3}(a) shows our results for $P^* \approx 5.29$ and $N$ varying from 22 to 76: in all depicted curves a series of allowed and forbidden $R$'s are clearly visible. As the size of the system increases, more and more of the allowed conformations are observed in both elongated --- $R>1$ and compressed --- $R<1$ shapes.
With the number of permitted conformations growing with $N$, one is faced with the problem of how to predict their structure. After examining numerous structures generated for numerous system sizes we came to the conclusion that their structure can be derived from the structure of the 2D disk system in the high density limit --- the triangular lattice \cite{Qi15}, at least at the qualitative level.  The derivation is carried out separately for nanoplatelets and nanorods. 

{\it Nanorods} are assembled by stacking  several columns of particles together along the long axis of the prism. If one column contains $n$ particles, then a column next to it will have $n-1$ particles, see  figure~\ref{fig3}(b) for an illustration. The columns are aligned so as to make a triangular lattice. If $m+1$ is the number of columns on each side facet, the total number of particles enclosed on all side facets is $N_{\text{side}}=(4n-2)m$. Index $m$ can be used to uniquely identify the type of the resulting structure. The rods with $m=1$ consist only of two columns. Consequently, the corresponding structure, $\textbf{R}_{1}$, has the largest possible aspect ratio. Structures with $m=2,3 \ldots$ have three columns and so on, giving rise to a decreasing series of aspect ratios, $\textbf{R}_{2}$,  $\textbf{R}_{3}$ and so on.  Particles on the top and bottom facets cannot make a triangular lattice due to geometrical considerations. We estimate their number using free volume as the guiding parameter. Specifically, it follows from figure~\ref{fig3}(b) that there can be no top particles in $\textbf{R}_{1}$ structures, only one particle in $\textbf{R}_{2}$ structures and two particles in $\textbf{R}_{3}$ structures. If we assume that the number of particles thus defined is a linear function of $m$ (a reasonable approximation for small $m$), we find that it must be $N_{\text{top}}=m-1$ in order to reproduce the estimates for $m=0,1$ and 2. Thus, the total number of particles confined to the surface of a prism with structure $\textbf{R}_{m}$ is $N=N_{\text{side}}+2N_{\text{top}}=4nm-2$, leading to an expression for the number of particles in one column  $n=\frac{N+2}{4m}$. This number can be invoked in the expression of the aspect ratio $L_{z}/L_{x}=\frac{2}{\sqrt{3}}\frac{N+2-4m}{4m^{2}}$, obtained after making substitutions $L_{x}=\frac{\sqrt{3}}{2} \sigma m$ and $L_{z}= \sigma (n-1)$,  leading to the final formula $R_{m}^{R}=\frac{2}{\sqrt{3}} \frac{N+2-4m}{4m^{2}}$, which describes how the aspect ratio of structure $\textbf{R}_{m}$ depends on the total number of particles in the system $N$. Figure~\ref{fig3}(b) demonstrates how well this dependence works for $\textbf{R}_{1}$, $\textbf{R}_{2}$ and $\textbf{R}_{3}$ by comparing theoretical predictions with simulation data. Overall there is a very good agreement between the two sets with average deviation not exceeding  3\% for $\textbf{R}_{1}$ and $\textbf{R}_{2}$ and 5\% for $\textbf{R}_{3}$, suggesting that the assumptions made in the derivation of the theoretical model are justified.

Nanoparticles with {\it platelet} geometry are discussed in figure~\ref{fig3}(c). Notations $\textbf{P}_{m}$ are used to distinguish these structures from nanorods. The platelet structure with the lowest $R$ is $\textbf{P}_{1}$, a direct analog of $\textbf{R}_{1}$. It consists of two planes covered with hard-sphere particles in a triangular lattice conformation superimposed on top of each other. To compute the aspect ratio of this structure, we first estimated the number of particles that fit within a square with the side comprising exactly $n$ particles, see figure~\ref{fig3}(c) for a graphical explanation. It turns out that this number depends on the scale of the square and for $2 \leqslant n \leqslant 7$ is well approximated by an expression $N_{\text{top}}=n(n-\frac{1}{2})$, which is suitable for system sizes up to 98 particles. The total number of particles in $\textbf{P}_{1}$ is  $N=2N_{\text{top}}=n(2n-1)$, leading to  $n=\frac{1}{4}+\frac{1}{2}\sqrt{\frac{1}{4}+2N}$. From the dimensions of the cell $L_{z}=\frac{\sqrt{3}}{2} \sigma$ and $L_{x}= \sigma (n-1)$ one then finds the aspect ratio $R_{1}^{P}=\frac{\sqrt{3}}{\sqrt{2N+0.25}-1.5}$.  Figure~\ref{fig3}(c) demonstrates how well this formula works for the studied systems. The deviation between the theoretical prediction and simulation is less than 8\% on average, which is small by all measures.  

Platelets with $m \geqslant 2$ are built similarly to rods except that now the columns are not aligned along the vertical axis but along the horizontal axis. Structures that have complete $m+1$ rows with either $n$ or $n-1$ particles in each of them, see figure~\ref{fig3}(c), form closed-off shells which can accept only new particles on the top or the bottom facets. Until these facets are full, the geometry of the cell does not change. Consequently, the aspect ratio of $\textbf{P}_{m}$ structures is predicted to vary in a step-wise manner as a function of the number of particles $N$. The exact values at which $R$ experiences a change can be estimated analytically.  It is seen from figure~\ref{fig3}(c) that a closed-off shell characterized by numbers $n$ and $m$  consists of  $m+1$ non-identical layers, each of which contains $2n+2(n-2)=4(n-1)$ particles. The total number of particles is $N=(m+1) \times 4 \times (n-1)$. For fixed $m$, this expression produces a series $N_{i}^{m}=4(m+1), 8(m+1), \ldots, 4i(m+1)$ at which $R(N)$ experiences a jump while remaining constant at intermediate points. The interval to which a given $N$ belongs can be identified by taking the integer part of the $\frac{N}{4(m+1)}$  ratio, i.e., computing $i= {\tt Int} \big(\frac{N}{4(m+1)}\big)$. Using this formula, the aspect ratio of platelets can be obtained from the cell geometry $L_{z}+\frac{\sqrt{3}}{2} \sigma m$ and $L_{x}= \sigma i$  as $R_{m}^{P}=\frac{\sqrt{3}}{2} \frac{m}{i}=\frac{\sqrt{3}}{2} \frac{m}{{\tt Int} \big(\frac{N}{4(m+1)}\big)}$, where $N$ is any number. How well this formula works for $\textbf{P}_{2}$ and $\textbf{P}_{3}$ is shown in figure~\ref{fig3}(c) plotting the theoretical and simulation results.  The step-wise character is clearly visible in the simulation data, in good qualitative agreement with theoretical predictions.  From the quantitative perspective, theoretical results are about 13\% accurate for $\textbf{P}_{2}$  structures and 21\% accurate for $\textbf{P}_{3}$  structures, on average. The agreement is worse than for the nanorods but still acceptable.

\begin{figure}[!t]
\centerline{\includegraphics[width=1\textwidth]{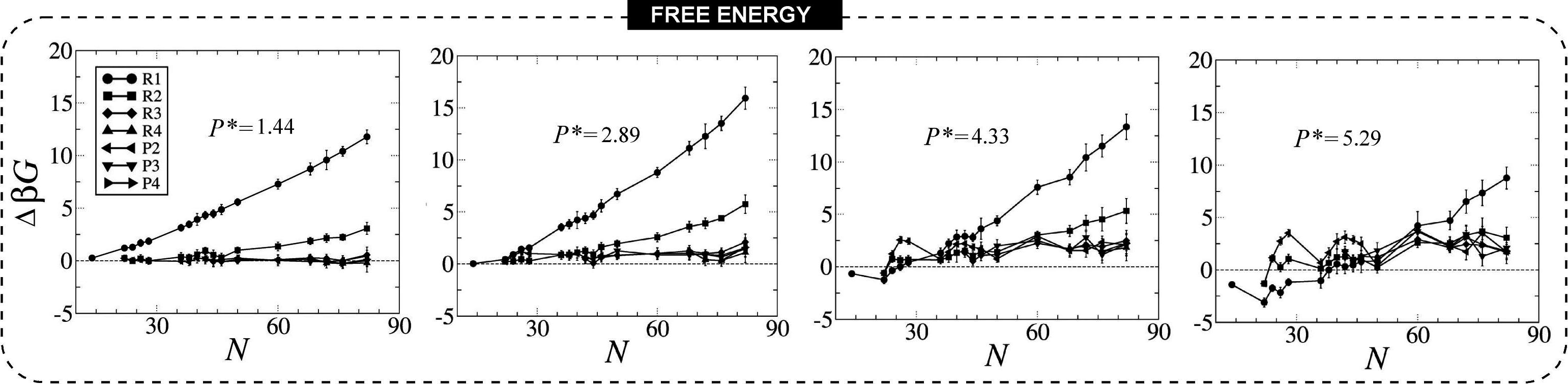}}
\caption{Relative free energy of different conformations as labeled with respect to $\textbf{P}_{1}$. Data are shown for different pressures as labeled. The number of particles is varied between 14 and 82.} 
\label{fig4}
\end{figure}

\subsection{Population of the allowed states}

Formulas obtained for $R_{m}^{P}$ and $R_{m}^{R}$ provide a means to determine the conformations that are permitted in a nanoparticle with $N$ ligands, both in the platelet and rod geometries. As a next step, we investigated the free energy of these conformations as a function of pressure. Our results are summarized in figure~\ref{fig4}, where free energy is shown relative to $\textbf{P}_{1}$ structure. The disparity between platelets and rods is clearly apparent. The rods are strongly destabilized as the number of particles increases. This trend is observed for $\textbf{R}_{2}$, $\textbf{R}_{3}$ and $\textbf{R}_{1}$, for which it is most prominent. As the pressure increases, the relative free energy $\Delta \beta G=\beta G(R_{i}) - \beta G(P_{1}), i=1, 2, 3$  also increases. At $P^{*}=1.44$ for instance,  $\Delta \beta G=12$ for $\textbf{R}_{1}$ but then it grows to $\Delta \beta G=15$ for $P^{*}=2.59$. Interestingly, the free energy difference begins to decline if the pressure increases beyond this point. At $P^{*}=5.29$ it becomes $\Delta \beta G=8$. This trend is observed for all rod structures but the magnitude of the free energy difference observed for $\textbf{R}_{2}$ and $\textbf{R}_{3}$ is much smaller than that of $\textbf{R}_{1}$.  At $P^{*}=5.29$, all rod structures have a free energy difference lower than $\Delta \beta G=5$.

The free energy difference of platelets is always less than $\Delta \beta G=5$, regardless of pressure or the number of particles used. Thus, it follows from our simulations that all identified conformations have approximately the same free energy. Elongated rod shapes with low ligand density on the surface are the exception to this rule.

\section{Discussion}

In systems interacting via hard-body potentials, the relevant thermodynamic function is entropy. Conformations observed experimentally or in simulations are those that exhibit the highest entropy among all alternative states. For instance, in the system of 2D disks on infinite plane --- the limiting model of the system considered here when $N \rightarrow \infty$ --- such conformations are the triangular lattice which emerges at considerably large densities, specifically at $\rho ^{*} >0.7$~\cite{Qi15}. It is a common knowledge that particles maximize the total volume in which they are allowed to move individually by participating in periodic collective movements of the lattice known as phonons. The large length scale needed to support phonons thus emerges as the key prerequisite for lattice conformations to be more stable than alternative amorphous-like states. 
The considered system is of finite size  and, as such, by definition cannot support periodic phonons. Yet, we see that triangular lattice conformations fully determine the geometry of the nanoparticle at high densities. Our simulations  show that particles placed in a specific structural motif  --- equilateral triangle ---  are capable of accessing a maximum conceivable volume, just as it happens in the case of  the long wavelength phonons. Local structure thus emerges as an important factor, along with the long-range order characteristic for lattices, contributing to the stability of the observed structure. This conclusion broadly agrees with the observation of the hexatic phase in 2D disks, which precedes the formation of the fully-ordered lattice conformations upon increasing density but possesses only a local structure, while making the studied system an excellent test-bed model for a general discussion over  the role of short {\it vs. } long order in phase transitions.

\begin{figure}[h]
	\centerline{\includegraphics[width=0.2\textwidth]{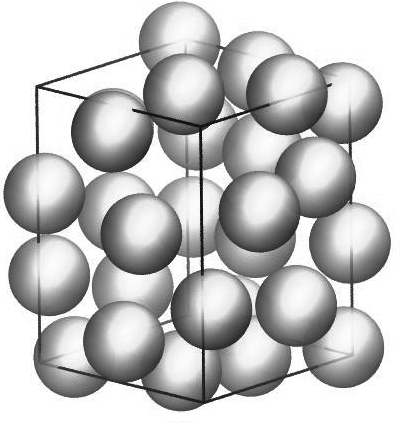}}
	\caption{A representative conformation of the system with 24 particles at $\rho ^{*} \approx 0.4$} 
	\label{fig5}
\end{figure}

As far as designing nanoparticles of particular shape is concerned, it is important to understand when specific shapes become predominant in the ensemble of all possible configurations. Of particular interest is to determine when the inter-conversion between different configurations becomes hampered, in other words at which point they become permanently locked. Our simulations show that for the system with $N=46$ particles, the free energy barriers between various conformations become greater than $5kT$ at $\rho ^{*} \approx 0.4$, see figure~\ref{fig2}(a) for evidence. This number varies a little with the size of the system. Thus, the configuration of the nanoparticle becomes determined much prior to the density at which structural transition into the solid state in the hard-sphere sub-system takes place, $\rho ^{*} \approx 0.7$. It is evident, therefore, that the hard-sphere sub-system defines the accessible nanoparticle configurations while remaining in the fluid state at a considerable degree of dilution. In support of this assertion, figure~\ref{fig5} depicts a representative conformation at $\rho ^{*} \approx 0.4$ in the system with $N=24$ particles. Furthermore, our simulations show that the particles experience no kinetic trapping at this density and can freely travel both within single facets as well as among different facets of the prism. Despite this, the hard-sphere fluid appears to populate certain specific shapes more frequently than others, which is quite surprising, given the common expectation that all shapes should be equally accessible to a system in a fluid state. It is not clear at the moment what mechanism underlies this phenomenon. What is clear is that ligands will exert a considerable influence over the shape of the emerging nanoparticle --- through excluded volume interactions --- even at a low surface coverage. Notably, ligands do not have to be arranged into any particular configurations in order to create strong structural preference. This effect should be taken into account when considering different choices of ligands that may vary in their size or shape.

\section{Acknowledgement}
The authors express their deepest gratitude to the men and women of the Ukrainian Armed Forces for making this work possible through their determination and selfless sacrifice while defending the independence and freedom of our country. AB acknowledges financial support from the Swiss National
Science Foundation 
(project no.~IZURZ$2$$\_224907$, 
Ukrainian-Swiss Joint Research Projects: Call for
Proposals 2023-UA-CH-NANO) and from the National Research Foundation of Ukraine (grant no.~2023.05/0019).

\newpage

 \ukrainianpart
 
 \title{Вплив виключеного об'єму лігандів на форму нанокристалу}
 \author{A. Баумкетнер\refaddr{label1}, Д. Анохін\refaddr{label2}, Я. Пацаган\refaddr{label3}}
 \addresses{
 	\addr{label1}
 	Iнститут фiзики конденсованих систем Нацiональної академiї наук України,  вул.~Свєнцiцького, 1,\\
 	79011 Львiв, Україна
 	\addr{label2}
 	Хімічний факультет, Харківський національний університет імені В.Н. Каразіна, майдан Свободи 4, \\61022 Харків, Україна
 	\addr{label3}
 	Фізичний факультет,
 	Львівський національний університет ім. Івана Франка,
 	вулиця Кирила і Мефодія, 8, 79005 Львів, Україна
 }

 \makeukrtitle
 
 \begin{abstract}
 Нами було досліджено вплив виключеного об'єму поверхнево-активних лігандів на форму квантових точок (КТ), до яких вони приєднані. Було розглянуто модель, в якій ліганди представлені твердими сферами, що зв'язані з поверхнею наночастинки (НК), що має форму призми. У моделюванні методом Монте-Карло виявлено, що ансамбль відповідних конформацій НК складається з невеликої кількості специфічних станів, які набувають форми нанопластин та наностержнів. Форму цих станів добре описують отримані теоретичні моделі. При збільшенні щільності лігандів вільна енергія різних станів приблизно однакова, що свідчить про те, що взаємодії виключеного об'єму між лігандами звужують конформаційний простір, доступний для НК, не створюючи статистичної переваги для будь-якої конкретної конфігурації.
 	\keywords нанокристали, тверді кульки, комп'ютерні симуляції
 	
 \end{abstract}
\end{document}